\documentclass[conference]{IEEEtran}
\IEEEoverridecommandlockouts
\usepackage{cite}
\usepackage{amsmath,amssymb,amsfonts}
\usepackage{algorithmic}
\usepackage{graphicx}
\usepackage{textcomp}
\usepackage{xcolor}

\def\BibTeX{{\rm B\kern-.05em{\sc i\kern-.025em b}\kern-.08em
    T\kern-.1667em\lower.7ex\hbox{E}\kern-.125emX}}
\begin{document}

\title{Focusing of a Long Relativistic Proton Bunch in Underdense Plasma\\
}

\author{\IEEEauthorblockN
{1\textsuperscript{st} Livio Verra}
\IEEEauthorblockA{
\textit{CERN}\\
Meyrin, Switzerland \\
livio.verra@cern.ch}
\and
{2\textsuperscript{nd} Edda Gschwendtner}\\
\IEEEauthorblockA{
\textit{CERN}\\
Meyrin, Switzerland \\
edda.gschwendtner@cern.ch}
\and
{3\textsuperscript{rd} Patric Muggli}\\
\IEEEauthorblockA{
\textit{Max Planck Institute for Physics}\\
Munich, Germany \\
muggli@mpp.mpg.de}
\and
{(AWAKE Collaboration)}

}

\maketitle

\begin{abstract}
We discuss the focusing effect experienced by a long relativistic proton bunch, when propagating in underdense plasma.  
We show with 2D quasi-static simulations that the response of the plasma to the presence of the bunch provides a focusing force for the protons. 
We discuss the impact of the finite transverse size of the plasma on the dynamics of the process and we introduce the measurements performed at the AWAKE experiment at CERN.
\end{abstract}

\begin{IEEEkeywords}
AWAKE, Proton-Driven Plasma Wakefield Acceleration, Self-Modulation, Plasma Focusing
\end{IEEEkeywords}

\section{Introduction}
When a relativistic charged particle bunch travels in a quasi-neutral plasma, the plasma electrons move, compensating for the essentially radial electric field $E_r$ of the bunch~\cite{PWFA:CHEN}.
Depending on the ratios between bunch charge density $n_b$ and the plasma electron density $n_{pe}$, between root mean square (rms) bunch length $\sigma_z$ and the plasma electron wavelength $\lambda_{pe}=2\pi c/\omega_{pe}$ ($\omega_{pe}=\sqrt{\frac{n_{pe} e^2}{m_e \varepsilon_0}}$ is the plasma electron angular frequency, where $e$ and $m_e$ are the electron charge and mass, $\varepsilon_0$ is the vacuum permittivity; $c$ is the speed of light), and between rms bunch radius $\sigma_r$ and the plasma skin depth $c/\omega_{pe}$, different modes of the beam-plasma interaction take place.
The propagation of a long and narrow ($\sigma_z=6\,$cm$\gg \sigma_r=0.15\,$mm) relativistic proton ($p^+$) bunch in a long and narrow ($L=10\,$m$\,\gg r_p\sim1\,$mm) plasma, whose plasma electron density can be varied over many orders of magnitude ($n_{pe}\approx10^{10} - 10^{15}\,$cm$^{-3}$) allows for the study of different beam-plasma interaction modes.

\par In the context of the Advanced Wakefield Experiment (AWAKE) at CERN~\cite{PATRIC:READINESS}, we so far focused on the self-modulation instability (SMI) mode: we demonstrated that, when $n_b<n_{pe}$, $\sigma_z\gg \lambda_{pe}$ and $\sigma_r\leq c/\omega_{pe}$, the bunch self-modulates in plasma~\cite{MARLENE:PRL, KARL:PRL}, and therefore resonantly drives large-amplitude wakefields that can be used for high-gradient particle acceleration~\cite{AW:NATURE}.

\par After propagation in preformed plasma, the transverse size of the bunch front, measured at a screen downstream of the plasma exit, is smaller than in the case without plasma, while in the back of the bunch the overall transverse size is larger due to the occurrence of SMI~\cite{LIVIO:PRL}.
The focusing effect associated with the smaller size of the bunch front was also observed for much lower values of $n_{pe}$~\cite{LIVIO:EPS22}, when the occurrence of SMI and the wakefield excitation are suppressed.
At very low density, the nature of focusing depends on the number of plasma electrons available to neutralize the bunch space-charge field, and thus on the plasma radius $r_p$.

\par We use numerical simulations to illustrate the type of interaction one can expect from the longitudinal and transverse fields excited by the $p^+$ bunch at the plasma entrance. 
We show that at very low plasma density ($n_b\gg n_{pe}$ and $\lambda_{pe}>\sigma_z$) the finite plasma radius $r_p\ll c/\omega_{pe}$ suppresses the excitation of wakefields and only plasma neutralization of the bunch charge occurs. 
The focusing force is proportional to the limited plasma electron charge available within $r_p$.
At higher density ($\lambda_{pe}<\sigma_z$), the bunch drives very low-amplitude wakefields that could develop into SMI.

\begin{figure*}[h!]
\centering
\includegraphics[scale=0.5]{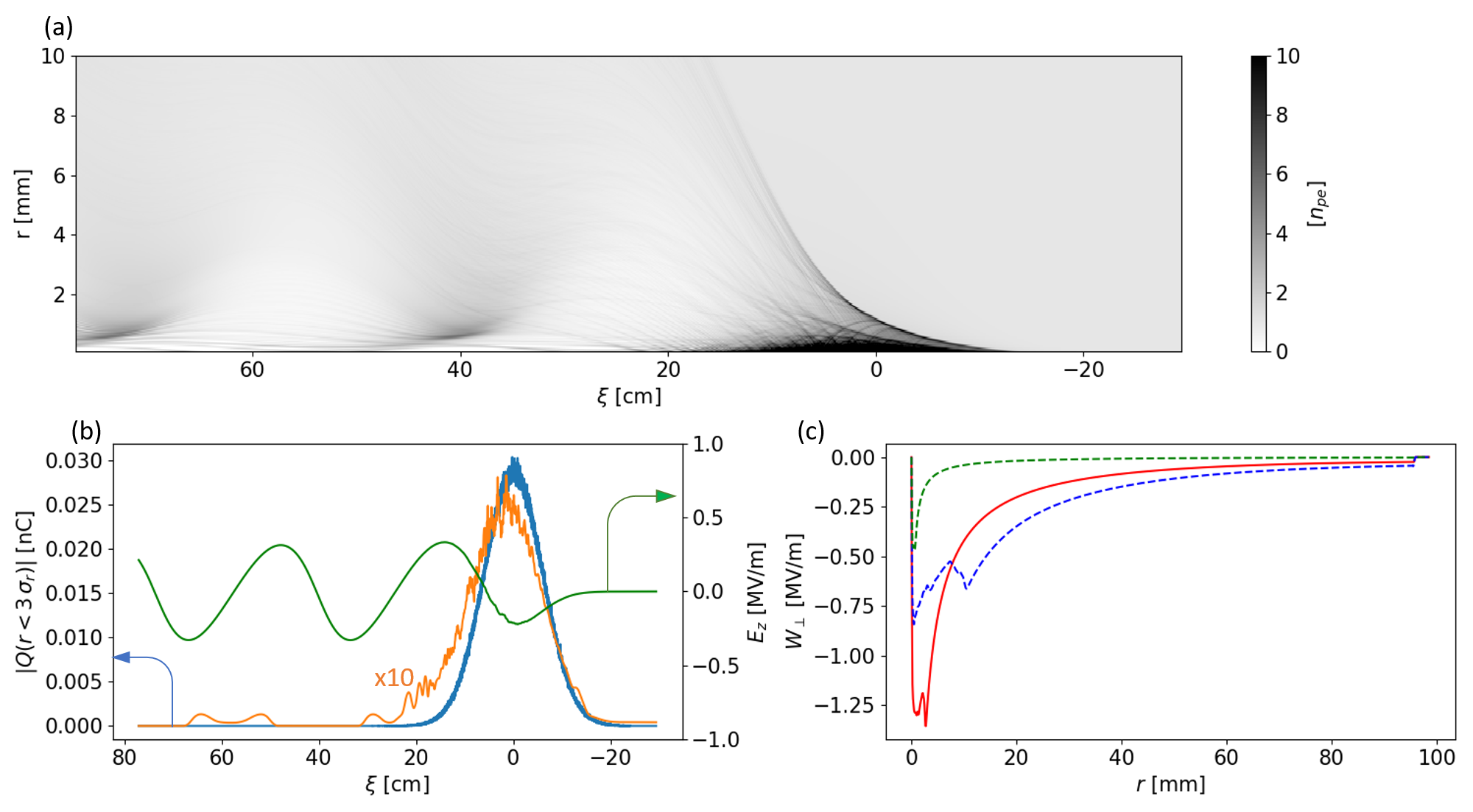}
\caption{
(a): Distribution of the plasma electron density in the cylindrical space ($\xi,r$) at $z=\lambda_{pe}$.
The bunch propagates from left to right.
Bunch center at $\xi=0$.
$n_{pe}=10^{10}\,$cm$^{-3}$, $r_p=1.8\times c/\omega_{pe}=95.76\,$mm.
(b): absolute charge of the $p^+$ bunch $Q_b$ (blue line, left hand-side vertical axis) and of the plasma electrons $Q_e$ (orange line multiplied by ten,left hand-side vertical axis) contained within $r<3\,\sigma_r$, and longitudinal electric field $E_z$ (green line, right hand-side vertical axis), as a function of $\xi$.
(c): transverse wakefields $W_{\perp}$ as a function of transverse radial position $r$, for $\xi=0$ (red solid line), $\xi=-1.5\,\sigma_z=-9\,$cm (green dashed line), and $\xi=+1.5\,\sigma_z=9\,$cm (blue dashed line).
Cell size is $1.7\times 10^{-3}\,k_{pe}^{-1}$ along $\xi$ and $1.9\times 10^{-3}\,k_{pe}^{-1}$ along $r$.
}
\label{fig:1}
\end{figure*}
\section{Simulation Results}
To investigate the modes of interaction accessible in the experiment, we perform particle-in-cell simulations with the 2D quasi-static code LCODE~\cite{LOTOV:LCODE,SOSEDKIN2016350:LCODE}.
To reproduce the experimental conditions, we simulate a $400\,$GeV/c $p^+$ bunch, with total charge $Q=44.8\,$nC, $\sigma_r=0.15\,$mm and $\sigma_z=6\,$cm ($n_b=1.3\cdot 10^{13}\,$cm$^{-3}$), with normalized transverse emittance $\epsilon_N=2.2\,$mm-mrad, propagating through initially uniform plasma.
Unlike in the experiment, in simulations we can vary the plasma radius to highlight its role in the mode of interaction. 

\par For $n_{pe}=10^{10}\,$cm$^{-3}\ll n_b$, the beam-plasma system is in the non-linear regime~\cite{ROSENZWEIG:NONLINEAR}, and $\lambda_{pe}>\sigma_z$.
In the case of simulations with plasma radius $r_p=1.8\times c/\omega_{pe} = 95.76\,$mm, the response of the plasma is similar to that of an infinitely wide plasma~\cite{FANG:doi:10.1063/1.4872328}, and there are enough plasma electrons to compensate for the space-charge field of the bunch. 
The distribution of the plasma electron density (Fig.~\ref{fig:1}(a)) in the cylindrical coordinates ($\xi,r$), after the bunch propagates for $z=\lambda_{pe}$, shows that the plasma electrons move towards the beam axis and that there the local plasma electron density reaches values more than ten times larger than $n_{pe}$.
Figure~\ref{fig:1}(b) shows that the distribution of charge of plasma electrons contained within $r< 3\,\sigma_r$ (orange line, multiplied by 10) essentially follows, but remains smaller than, that of the $p^+$ bunch (blue line). 
The oscillation of the longitudinal electric field $E_z$ behind the bunch with wavelength close to $\lambda_{pe}$ indicates that the bunch drives wakefields.
The transverse wakefields $W_{\perp}$ (Fig.~\ref{fig:1}(c), calculated as the sum of the radial electric and azimuthal magnetic fields), which peak and start decaying after $r\sim 0.5\,c/\omega_{pe}$, are focusing for the $p^+$ bunch. 
The amplitude depends on the longitudinal position along the bunch: when evaluated at the bunch front ($\xi=-1.5\,\sigma_z=-9\,$cm, green dashed line), it is smaller than at the back ($\xi=+1.5\,\sigma_z=+9\,$cm, blue dashed line), and it is maximum at the bunch center ($\xi=0$, red solid line).

\par
\begin{figure*}[h!]
\centering
\includegraphics[scale=0.5]{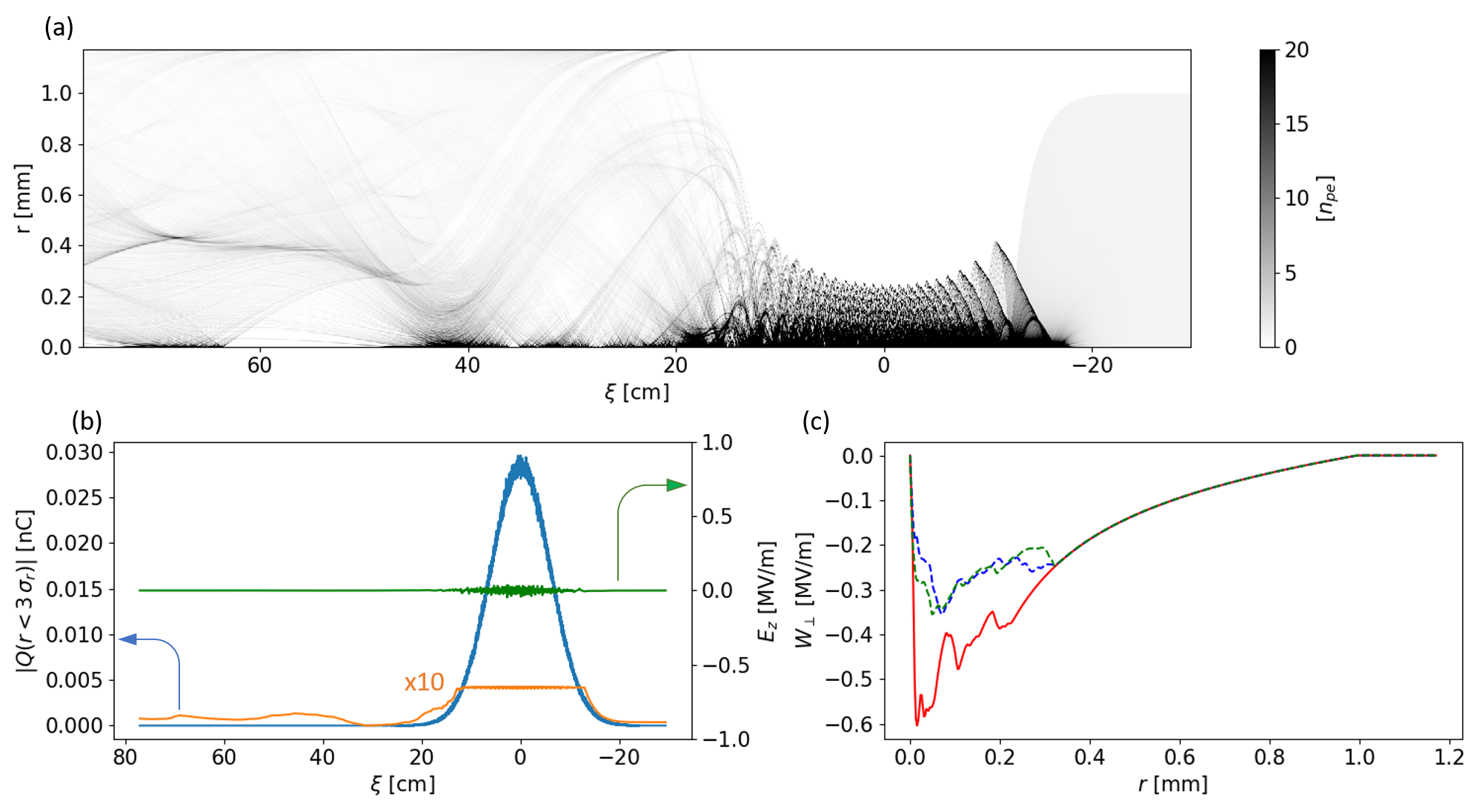}
\caption{(a): Distribution of the plasma electron density in the cylindrical space ($\xi,r$) at $z=\lambda_{pe}$.
The bunch propagates from left to right.
Bunch center at $\xi=0$.
$n_{pe}=10^{10}\,$cm$^{-3}$, $r_p=1\,$mm as in the experiment.
(b): absolute charge of the $p^+$ bunch $Q_b$ (blue line, left hand-side vertical axis) and of the plasma electrons $Q_e$ (orange line multiplied by ten,left hand-side vertical axis) contained within $r<3\,\sigma_r$, and longitudinal electric field $E_z$ (green line, right hand-side vertical axis), as a function of $\xi$.
(c): transverse wakefields $W_{\perp}$ as a function of transverse radial position $r$, for $\xi=0$ (red solid line), $\xi=-1.5\,\sigma_z=-9\,$cm (green dashed line), and $\xi=+1.5\,\sigma_z=9\,$cm (blue dashed line).
Cell size is $1.7\times 10^{-3}\,k_{pe}^{-1}$ along $\xi$ and $4.9\times 10^{-5}\,k_{pe}^{-1}$ along $r$.
}
\label{fig:2}
\end{figure*}

 To understand the effect of the transverse extent of the plasma, we also run a simulation with $r_p=1\,$mm$\ll c/\omega_{pe}$, as in the experiment~\cite{GABOR:PLASMA}.
Figure~\ref{fig:2}(a) shows that the result is very different from the previous case (compare to Figure~\ref{fig:1}(a)). 
In each longitudinal slice along the bunch, all the available plasma electrons are sucked in towards the beam center, where they generate a high-density ($>20\,n_{pe}$) filament close to the propagation axis~\cite{PABLO:FILAMENT,JAN:FILAMENT}.
For example at $\xi = 0$, $n_{pe}=0$ for $r>0.4\,$mm.
This is also shown by the charge distribution of the plasma electrons contained within $r< 3\,\sigma_r$ (Fig.~\ref{fig:2}(b), orange line), that remains constant along most part of the bunch, and that is much smaller than in the previous case, because of the smaller number of available electrons.
Moreover, some electrons move so abruptly that they overshoot, generating caustic arcs (Fig.~\ref{fig:2}(a))~\cite{PABLO:FILAMENT,HOGAN:POSITRONS}.
In this case, due to $r_p\ll c/\omega_{pe}$, the plasma electrons do not react collectively to the presence of the bunch, as in a wave, and therefore the wakefields are suppressed. 

\par The amplitude of the longitudinal electric field $E_z$ is zero along and behind the bunch.
The transverse fields $W_{\perp}$ at $\xi=0$ (Fig.~\ref{fig:2}(c), red solid line), focusing for protons, spike at $r\sim 0.01\,$mm$\ll \sigma_r$, and decay with a $1/r$ dependency.
This indicates again that the plasma electrons concentrate very close to the beam axis.
At $\xi=\pm 1.5 \sigma_z$, the amplitude of the transverse fields remains approximately constant until $r\sim 0.4\,$mm and then it starts decaying, following the same curve as at $\xi=0$.
This shows that the same amount of plasma electrons (all the available ones) move towards the axis and are contained within $r<0.4\,$mm.
It also means that the amplitude depends on the local bunch charge density, and that the dynamics of each longitudinal slice does not depend on the evolution of the preceding ones.

\par When $r_p\ll c/\omega_{pe}$ as in this case, increasing the number of available plasma electrons while keeping $r_p$ fixed (i.e., increasing $n_{pe}$, as in the experiment), translates into a stronger focusing effect. 

\par For both plasma radii, $W_{\perp}$ does not depend linearly on $r$ (Fig.s~\ref{fig:1}(c) and \ref{fig:2}(c)), which means that the normalized transverse emittance of the bunch is not preserved upon propagation in plasma.
\par
\begin{figure*}[h!]
\centering
\includegraphics[scale=0.5]{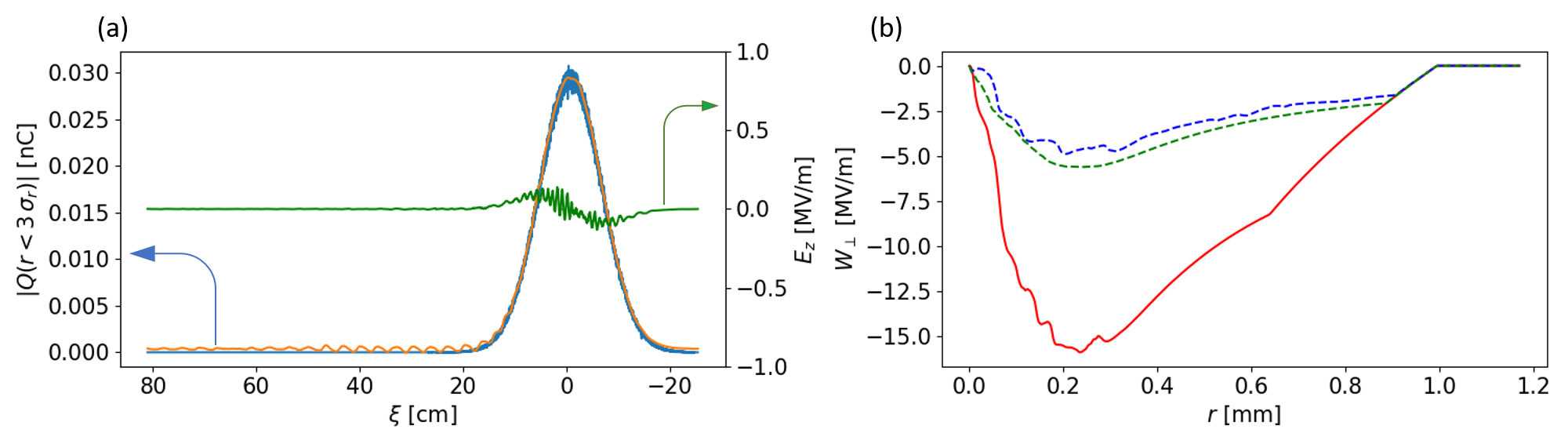}
\caption{
(a): absolute charge of the $p^+$ bunch $Q_b$ (blue line, left hand-side vertical axis) and of the plasma electrons $Q_e$ (orange line,left hand-side vertical axis) contained within $r<3\,\sigma_r$, and longitudinal electric field $E_z$ (green line, right hand-side vertical axis), as a function of $\xi$.
(b): transverse wakefields $W_{\perp}$ as a function of transverse radial position $r$, for $\xi=0$ (red solid line), $\xi=-1.5\,\sigma_z=-9\,$cm (green dashed line), and $\xi=+1.5\,\sigma_z=9\,$cm (blue dashed line).
Cell size is $1.7\times 10^{-2}\,k_{pe}^{-1}$ along $\xi$ and $4.9\times 10^{-4}\,k_{pe}^{-1}$ along $r$.
$n_{pe}=10^{12}\,$cm$^{-3}$, $r_p=1\,$mm.}
\label{fig:3}
\end{figure*}

 Figure~\ref{fig:3}(a) shows that, for $n_{pe}=10^{12}\,$cm$^{-3}<n_b$ and ${r_p=1\,}$mm$<c/\omega_{pe}$, the plasma electron distribution (orange line) mimics that of the $p^+$ bunch (blue line).
The low-amplitude oscillation of the plasma electron distribution with wavelength close to $\lambda_{pe}$ behind the bunch ($\xi>20\,$cm) indicates that the bunch drives wakefields (not visible in $E_z$).
The non-uniformity of the plasma electron distribution generates a longitudinal electric (green line) as well as transverse (Fig.~\ref{fig:3}(b)) focusing fields along the bunch that dominate the beam-plasma interaction.
The amplitude of $W_{\perp}$ is larger than in the lower $n_{pe}$ cases (Fig.s~\ref{fig:1}(c) and ~\ref{fig:2}(c)), because the plasma electrons fully compensate for the space-charge field of the bunch.
In this case, $\lambda_{pe}=3.3\,$cm$<\sigma_z$.
Therefore, SMI could grow and become observable after propagation in a long plasma.

\section{Experimental setup and measurements}

In AWAKE, the plasma is generated by ionizing rubidium vapor with a $\sim120\,$fs, $\sim100\,$mJ laser pulse. 
Previous experiments~\cite{KARL:PRL} showed that the pulse ionizes $\sim100\,\%$ of the vapor along its path, producing a $\sim 1$-mm-radius plasma column with density equal to that of the vapor.
The vapor is provided by a $10$-m-long vapor source~\cite{PATRIC:READINESS,GENADY:SOURCE}. 
Rubidium is contained in two reservoirs located at each end of the source.
The vapor flows from the reservoirs to the source.
The temperature of the reservoirs $T$ defines the vapor pressure of the rubidium they contain.
When increasing $T$, the vapor flow increases, increasing the vapor density in the source. 
\par The $400\,$GeV/c $p^+$ bunch is delivered by the CERN Super Proton Synchrotron (SPS). 
The bunch is focused at the plasma entrance with the parameters listed in the previous section.
Downstream of the vapor source, the ionizing laser pulse is stopped by a beam dump.
An aluminum-coated silicon wafer generates optical transition radiation (OTR) when the $p^+$ bunch goes through it. 
The OTR is transported and imaged onto the entrance slit of a streak camera, that produces time-resolved images of the $p^+$ bunch charge density distribution~\cite{KARL:STREAK}.
Since the screen is located $3.5\,$m downstream of the plasma exit, the effect of propagation in vacuum is also convoluted into the bunch distribution observed at the screen.
Moreover, the transverse emittance of the bunch is likely to increase while propagating in plasma, because the focusing field is not linear with radius, as shown by the simulation results (Fig.s~\ref{fig:1}(c) and~\ref{fig:2}(c)).
\par For studying the focusing of the $p^+$ bunch, we measure the longitudinal charge density distribution, after propagation in plasma, while varying $n_{pe}$.
Increasing $n_{pe}$ from $10^{10}$ to $10^{15}\,$cm$^{-3}$, we investigate all the beam-plasma interaction modes mentioned above.
For very low densities, SMI does not take place and the transverse size of the beam at the screen is smaller than in the case without plasma, due to the focusing effect.
When increasing the density ($r_p\ll c/\omega_{pe}$), the measured transverse size of the beam decreases, as an increasing number of plasma electrons are available in the plasma.
Since the bunch does not drive wakefields, each longitudinal slice evolves according to its own charge density and does not affect the evolution of the following ones~\cite{LIVIO:PRL}.
When increasing $n_{pe}$ further, the focusing effect reaches a maximum, and the transverse size measured at the screen remains constant until SMI becomes the dominant effect~\cite{LIVIO:PRL}.
We will present these experimental results in a dedicated publication~\cite{LIVIO:FOCUSING}.

\section{Conclusions}

In the context of the AWAKE experiment, we investigate the propagation and focusing of a long relativistic $p^+$ bunch in plasma with a broad range of plasma electron densities.
The response of the plasma to the presence of the bunch focuses the bunch.
We show with 2D quasi-static particle-in-cell simulations that, at very low densities, the strength of the focusing force increases with the number of available plasma electrons, until it reaches saturation when the space-charge field of the bunch is fully compensated.

\par The bunch and plasma parameters at the plasma entrance determine the type of interaction: at very low densities (${n_{pe}\ll n_b}$, $\sigma_z>\lambda_{pe}$), non-linear wakefields occur for $r_p\gg c/\omega_{pe}$, while partial compensation and wakefields suppression take place for $r_p\ll c/\omega_{pe}$.
For larger $n_{pe}$ ($\sigma_z<\lambda_{pe}$), the bunch drives low-amplitude wakefields, from which SMI may grow.
These interactions can be observed in experiments after the bunch has propagated along the 10-m-long plasma.


\bibliography{version3.bib}{}
\bibliographystyle{IEEEtran}

\end{document}